\documentclass[12pt,letterpaper,reqno]{amsart}

\usepackage{times} 
\usepackage[T1]{fontenc} 
\usepackage{mathrsfs} 
\usepackage{latexsym} 
\usepackage[dvips]{graphics}
\usepackage{epsfig}
\usepackage{amsmath,amsfonts,amsthm,amssymb,amscd} 
\input amssym.def 
\input amssym.tex

\newcommand\be{\begin{equation}} 
\newcommand\ee{\end{equation}}
\newcommand\bea{\begin{eqnarray}} 
\newcommand\eea{\end{eqnarray}} 
\newcommand\bi{\begin{itemize}}
\newcommand\ei{\end{itemize}} 
\newcommand\ben{\begin{enumerate}} 
\newcommand\een{\end{enumerate}}
\newcommand\bc{\begin{center}} 
\newcommand\ec{\end{center}} 
\newcommand\ba{\begin{array}} 
\newcommand\ea{\end{array}}

\theoremstyle{definition} 

\begin{document}

\title[Balance in social networks] 
{On the notion of balance in social network analysis}


\author{Peter Hegarty} \address{Department of Mathematical Sciences, 
Chalmers University Of Technology and University of Gothenburg,
41296 Gothenburg, Sweden} \email{hegarty@chalmers.se}


\date{\today}

\begin{abstract} 
The notion of ``balance'' is fundamental for sociologists who study social networks. In formal mathematical terms, it concerns the distribution of triad configurations in actual networks compared to random networks of the same edge density. On reading Charles Kadushin's recent book ``Understanding Social Networks'', we were struck by the amount of confusion in the presentation of this concept in the early sections of the book. This confusion seems to lie behind his flawed analysis of a classical empirical data set, namely the karate club graph of Zachary. Our goal here is twofold. Firstly, we present the notion of balance in terms which are logically consistent, but also consistent with the way sociologists use the term. The main message is that the notion can only be meaningfully applied to undirected graphs. Secondly, we correct the analysis of triads in the karate club graph. This results in the interesting observation that the graph is, in a precise sense, quite ``unbalanced''. We show that this lack of balance is characteristic of a wide class of starlike-graphs, and discuss possible sociological interpretations of this fact, which may be useful in many other situations.      
\end{abstract}


\maketitle

\setcounter{equation}{0}

\setcounter{equation}{0}

\setcounter{section}{0}

\section{Introduction}

Social Network Analysis, henceforth abbreviated to SNA, is an area of research 
which has seen an explosion of activity in recent years, 
with a flood of both academic 
research papers and more popular literature. The field is a 
paradigm of ``cross-disciplinary research'', attracting the attention of people from a wide range of academic specialisations. The 
opposite ends of this spectrum of specialisations are essentially occupied by 
sociologists and mathematicians. Sociologists often do the groundwork of 
collecting empirical data and compiling them into networks. This work is 
crucial - without it, no scientific analysis is possible and the field ceases 
to exist. Quantitative analysis of social networks often involves the
comparison of real networks with randomly generated ones, and the search for 
patterns in the actual networks which occur with a frequency far different 
from what one would expect if links were formed completely at random. Such 
comparative analysis can be mathematically quite sophisticated, and in general 
requires the analyst to have a good working knowledge of that branch of 
discrete mathematics known as ``random graphs''. 
\par I am a mathematican with a background in discrete mathematics, who has been recently taking part in a reading 
course on SNA (see the acknowledgement below) out of simple curiosity about 
this exciting area. The participants in this course reflect, in the best 
possible manner, the interdisciplinary nature of the field, and several of the 
texts we have been using are written primarily for an audience of sociologists 
with limited mathematical training. One of these is a recently published text 
by Charles Kadushin \cite{Ka}, a major figure on the sociological 
side of SNA. As it states on the back cover, the book is ``{\em aiming for 
those interested in this fast-moving area who are not mathematically 
inclined}''. Nevertheless, the book does employ some mathematical terminology 
and present some explicitly quantitative analyses. Such effort can in general 
only be applauded, and a mathematician should approach such a text in a spirit 
of generosity. However, I quickly uncovered problems with this book of a very 
serious nature. Fundamental concepts, both sociological and mathematical, are introduced in a way which simply does not make sense. The first quantitative analysis of an actual network, the celebrated karate club network of Zachary \cite{Z}, is deeply flawed.  
\par It's not my 
purpose here to do a comprehensive book review - all the problems I will 
discuss arise, after a general introductory chapter, in the first 17 pages of 
the substantive text. Rather I want to correct the author's 
presentation of some 
fundamental concepts in a way which might prove useful to researchers and 
students in the future, especially to sociologists who
might be interested in seeing how a mathematician approaches this material. 
I shall be primarily concerned with the mathematical 
notion of {\em transitivity} and its application to the sociological notion of
the same name, along with the more restrictive notion of  
{\em balance}. I shall discuss these terms in a manner which is logically 
consistent, but also consistent with the way sociologists try to apply them. 
In doing so, I will explain what is wrong with Kadushin's text, the crucial 
point being that the concept of balance cannot be meaningfully discussed for 
graphs unless they are undirected. This material is presented in Section 3. 
\par In Section 4 we perform a correct triad census for the karate club graph 
of Zachary, which involves comparison of the actual counts of different triad 
configurations with those in an Erd\H{o}s-Renyi random graph of the same 
(expected) edge 
density. Though the 
mathematics involved is ``standard'', I will present it in detail. The 
presentation of this material in the book 
is deeply flawed, as the author compares the actual 
network with random {\em directed} graphs. He is led to the qualitatively 
false conclusion that Zachary's graph is very balanced. The correct 
analysis leads to a quite different, and more interesting conclusion. In 
Zachary's graph, triads with one edge out of three present are significantly 
underrepresented, compared to corresponding random graphs, whereas all other
triad configurations are overrepresented. The graph is therefore quite 
unbalanced. 
\par In Section 5, I show that the distribution of triads observed in 
Zachary's graph is characteristic of a precisely defined class of ``starlike'' 
networks. This is the mathematically most demanding part of the article. A
reader not primarily interested in rigorous proofs may therefore choose
to just skim over Section 5 and jump ahead to Section 6, where I discuss what 
I think are plausible sociological interpretations of such networks, and of
unbalanced networks in general, and their
relevance to understanding the social dynamics in Zachary's karate club. 
\par In Section 7, I will revisit the concept of balance itself. 
On the one hand, I will
show that, with a small change in the basic definitions, balance automatically
incorporates dyadic symmetry, something which might help avoid the 
kind of confusion which arose in \cite{Ka}. On the other hand, I will
discuss what seems to be the obvious notion of ``balance'' which makes sense 
for any weighted digraph.
The quotation marks here are important, because the notion I propose
is quite different from that which is used in 
sociology, so much so that 
a new term would need to be invented for it.  
\par Section 8 is a short discussion of some inevitably controversial issues 
which this note raises. 

\setcounter{equation}{0}

\section{Graph notation and terminology}

The following notation and terminology is standard, but it is important that 
we leave no room for doubt as to what statements in subsequent sections mean. 
Non-mathematicians may also find this section useful. A {\em 
directed graph (digraph)} is a pair $(V,E)$,
where $V$ is a finite set of so-called {\em nodes}, and $E$ is a set of 
ordered pairs $(v_1,v_2)$, where $v_1$ and $v_2$ are distinct elements of 
$V$. The ordered pair $(v_1,v_2)$ is referred to as the {\em directed edge
from $v_1$ to $v_2$}, and written symbolically as $v_1 \rightarrow v_2$.
Note that our definition allows for the existence of up to two 
directed edges between a given pair of nodes, one in each direction. We
disallow {\em loops}, i.e.: edges from a node to itself, though one should
keep in mind that, in many
social networks, it is implicit in the meaning of the edges that a
loop exists at each node. 
\par Given a digraph $G = (V,E)$, and a subset $V^{\prime} \subseteq V$, we can
consider the digraph $H = (V^{\prime}, E^{\prime})$ whose nodes are the 
elements of $V^{\prime}$ and whose edge-set $E^{\prime}$ consists of those
directed edges $v_1 \rightarrow v_2$ in $E$ such that both $v_1$ and $v_2$ lie
in $V^{\prime}$. We refer to $H$ as the {\em sub(di)graph} of $G$ {\em induced} on
the subset $V^{\prime}$. Of particular importance in this paper will
be subgraphs induced on 2 or 3 nodes. A digraph on 2 nodes is called a 
{\em dyad}, while one on 3 nodes is called a {\em triad}{\footnote{The 
terminology of dyads and triads is used more by sociologists than 
mathematicians.}}. 
\par A digraph is said to be {\em symmetric} if, for each pair $v_1, v_2$ of
distinct nodes, the directed edges $v_1 \rightarrow v_2$ and 
$v_2 \rightarrow v_1$ are either both present or both absent. The
description of such digraphs can be simplified by replacing
each existing pair of directed edges by a single undirected edge. This yields 
what we shall simply call a {\em graph}, i.e.: the word ``graph'' on its own
means that the edges are undirected. We shall also use the terms ``dyad''
and ``triad'' for graphs on 2 and 3 nodes respectively, and it will
always be clear from the context whether we are considering
graphs or digraphs. 
\par For graphs it is clear that there are only 
two possible dyads, since a single edge is either present or not. Given three
nodes $A,B$ and $C$, there are $2^3 = 8$ possibilities for a graph on these
three nodes, since each of 3 possible edges can be present or not. However,
these 8 graphs fall into only four {\em isomorphism classes} or {\em 
configurations}, the latter being the term of choice for sociologists. 
In general, two graphs (resp. digraphs) are said to be {\em isomorphic} if
they contain exactly the same edges (resp. directed edges) up to a 
relabelling of the nodes. For graph triads, the isomorphism class is
completely determined by the number of edges present{\footnote{This is not
true for larger numbers of nodes. Indeed, it is a very difficult problem
to determine the number of isomorphism classes of graphs on $n$ nodes, when
$n$ is large. See \cite{O}.}}, which can be 0,1,2 or 3. 
So, for example, given nodes $A,B,C$, the graph containing only the 
edge between $A$ and $B$ is isomorphic to that containing the single edge
between $B$ and $C$, since the latter graph can be got from the former
by relabelling the nodes $A,B,C$ as $C,B,A$ respectively. Of a total of 8 
possible graphs, there are 1,3,3 resp. 1 in the isomorphism classes 
with 0,1,2 resp. 3 edges.  
Finally, note that 
a graph on 3 nodes with all 3 edges present is usually called a {\em triangle},
whereas one where no edges are present is said to be {\em empty}. If 
exactly 2 edges are present, the triad is called {\em intransitive} 
(see Section 3 below). 
\par For digraphs, there are 3 isomorphism classes of 
dyads, depending on whether neither, exactly one of, or both the two
possible directed edges are present. 
It is a more challenging exercise to verify that there are 16 
isomorphism classes of digraph triads. This fact is well-known to 
sociologists, however, who have also adopted a conventional numbering of
the 16 possibilities. The complete list of digraph triads can be found on 
page 24 of \cite{Ka}, along with the conventional numbering. It's important to 
keep in mind that, given three
nodes $A,B,C$, there are $2^6 = 64$ possibilites for a digraph on these three
nodes, since each of 6 possible directed edges can be present or not. 
However, the 64 digraphs fall into just 16 isomorphism classes. With 
respect to the
conventional numbering, it can be checked that the number of digraphs in 
each class is given by the sequence of 16 numbers
\begin{equation}
1,6,3,3,3,6,6,6,6,2,3,3,3,6,6,1.
\end{equation}

\setcounter{equation}{0}

\section{Transitivity and balance}

Transitivity is a basic concept with a precise meaning in mathematics. In SNA, 
the notion is captured informally with the motto
\\
\\
M1. ``the friend of my friend is my friend''.
\\
\\
To make this motto precise, we may consider a digraph, where the nodes 
represent people, and where a directed edge from $x$ to $y$ means that $x$ 
considers $y$ as his/her friend. Then a formal statement of M1 is the 
following:
\\
\\
M1. Let $x,y,z$ be three distinct nodes in a digraph. If the directed edges 
$x \rightarrow y$ and $y \rightarrow z$ are both present, then so is the 
directed edge $x \rightarrow z$. 
\\
\\
This is very close to the formal definition of transitivity in mathematics, 
the only difference being that, in the latter, the nodes $x,y$ and $z$ are not 
assumed to be distinct. In sociology, the notion of transitivity leads 
naturally to that of {\em balance}. The latter is captured informally by M1 
along with three further, similar-sounding mottos:
\\
\\
M2. ``the enemy of my enemy is my friend''
\\
M3. ``the enemy of my friend is my enemy''.
\\
M4. ``the friend of my enemy is my enemy''.
\\
\\
The corresponding formal statements are then as follows:
\\
\\
M2. Let $x,y,z$ be three distinct nodes in a digraph. If the directed edges 
$x \rightarrow y$ and $y \rightarrow z$ are both absent, then the directed edge
$x \rightarrow z$ is present.
\\
\\
M3. Let $x,y,z$ be three distinct nodes in a digraph. If the directed edge 
$x \rightarrow y$ is present and the directed edge $y \rightarrow z$ is absent,
then the directed edge $x \rightarrow z$ is absent.
\\
\\
M4.  Let $x,y,z$ be three distinct nodes in a digraph. If the directed edge 
$x \rightarrow y$ is absent and the directed edge $y \rightarrow z$ is present,
then the directed edge $x \rightarrow z$ is absent.
\\
\\
Formally, balance is a property of digraph triads. A digraph triad is
said to be {\em (completely) balanced} if M1-M4 all hold. It is a
straightforward but tedious exercise to verify that a balanced triad must be
symmetric, and the resulting graph must then contain either 1 or 3 edges. 
Indeed, the table on the next page shows 
which of the properties M1-M4 hold for each
of the 16 isomorphism classes of digraph triads (Y indicates that the 
property holds, N that it doesn't). Here is an example to assist the reader.

\begin{figure*}[ht!]
  \includegraphics[width=\textwidth]{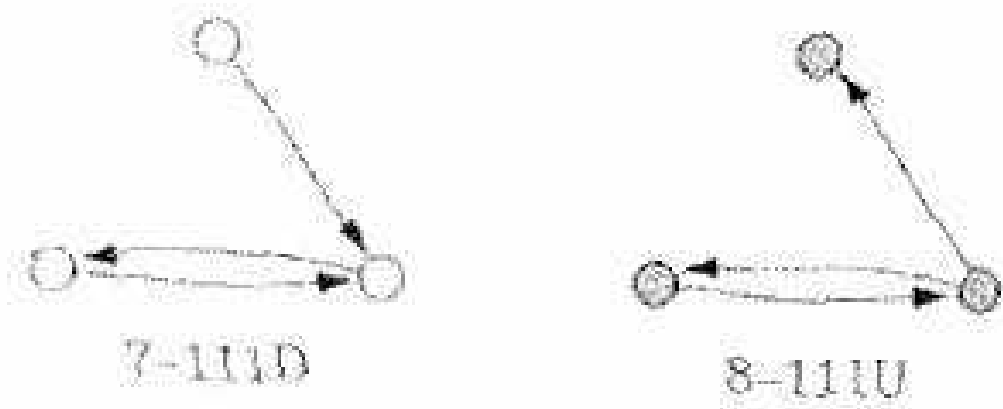} 
 \label{fig:ini}
\caption{Triad types 7 and 8, reproduced from page 24 of \cite{Ka}.}
\end{figure*}
 
Consider triad type 7, which is the graph on the left of Figure 1. Call 
the three vertices $A,B,C$, starting from the bottom left corner
and reading counter-clockwise. Hence this triad contains the three directed
edges $A \rightarrow B$, $B \rightarrow A$ and $C \rightarrow B$. The 
ordered triple $(C,B,A)$ fails to satisfy M1, since $C \rightarrow B$ and
$B \rightarrow A$ are both present, but $C \rightarrow A$ is absent. 
The triple $(A,C,B)$ fails to satisfy M4, since $A \rightarrow C$ is absent
whereas $C \rightarrow B$ and $A \rightarrow B$ are both present. The
triple $(C,A,B)$ also fails to satisfy M4. 

\begin{table}[ht!]
\begin{center}
\begin{tabular}{|c|c|c|c|c|} \hline
${\hbox{Triad type}}$ & ${\hbox{M1}}$ & ${\hbox{M2}}$ & {\hbox{M3}} & {\hbox{M4}} \\ \hline \hline
$1$ & $Y$ & $N$ & $Y$ & $Y$ \\ \hline
$2$ & $Y$ & $N$ & $Y$ & $Y$ \\ \hline
$3$ & $Y$ & $Y$ & $Y$ & $Y$ \\ \hline
$4$ & $Y$ & $N$ & $N$ & $Y$ \\ \hline
$5$ & $Y$ & $N$ & $Y$ & $N$ \\ \hline
$6$ & $N$ & $N$ & $Y$ & $Y$ \\ \hline
$7$ & $N$ & $Y$ & $Y$ & $N$ \\ \hline
$8$ & $N$ & $Y$ & $N$ & $Y$ \\ \hline
$9$ & $Y$ & $N$ & $N$ & $N$ \\ \hline
$10$ & $N$ & $Y$ & $Y$ & $Y$ \\ \hline
$11$ & $N$ & $Y$ & $N$ & $N$ \\ \hline
$12$ & $Y$ & $Y$ & $Y$ & $N$ \\ \hline
$13$ & $Y$ & $Y$ & $N$ & $Y$ \\ \hline
$14$ & $N$ & $Y$ & $N$ & $N$ \\ \hline
$15$ & $N$ & $Y$ & $N$ & $N$ \\ \hline
$16$ & $Y$ & $Y$ & $Y$ & $Y$ \\ \hline
\end{tabular} 
\end{center}
\end{table}
$\;$ \\
For the sociologist, a potential use of mottos M1-M4 is to 
make predictions about unseen parts of a social network. For example, suppose 
we have three people $A,B$ and $C$, and have only 
been able to observe directly the interactions between two pairs, $A$ and $B$,
respectively $B$ and $C$. Then based on our observations and the mottos
M1-M4, we could try to make predictions about the unobserved relationship
between $A$ and $C$.  
The fact that a balanced triad must be symmetric then assumes crucial 
importance, since it implies that, as a 
matter of pure logic, the mottos M1-M4 cannot make unambiguous predictions 
about unobserved social relationships, unless the observed relationships are 
symmetric{\footnote{Sociologists use the word {\em mutual} in this context.}}. 
\par To drive this crucial point home, we consider an example. Suppose we have 
a friendship network 
and three entities $A,B,C$. Suppose, for example, that $A$ and $B$ have been 
observed to like one 
another, whereas $B$ likes $C$, but $C$ dislikes $B$ (see triad type 8, to the
right in Figure 1). Hence, at least one 
pairwise relationship is not symmetric. However, we have full information about 
two pairs, so if the mottos M1-M4 are to be of any use in this situation, 
then it 
should be possible to make unambigous predictions about the relationships in 
the third pair. So we ask the question, should one expect $A$ to like $C$ or 
not, i.e.: should the directed edge $A \rightarrow C$ be present in the 
network ? 
Well, on the one hand, $A$ likes $B$ and $B$ likes $C$, so M1 suggests that, 
yes, $A$ should like $C$. But suppose $A$ does in fact like $C$. Then 
$A$ likes $C$, but 
$C$ dislikes $B$, so M3 suggests that $A$ should also dislike $B$. But 
$A$ likes $B$, a contradiction.     
\\
\\
In sociology, the first mention of the idea of balance is generally 
attributed to Heider. A direct citation from Heider's work 
appears on page 23 of \cite{Ka}:
\\
\par ``{\em In the case of three entities, a balanced state exists if all 
three relations are positive in all respects, or if two are negative and one 
is positive (Heider 1946, 110)}''.
\\
\\
In Heider's formulation it is clear that ``balance'' is to be considered as a 
property of the collection of pairwise relationships between three entities, 
in which each pairwise relationship is already mutual (positive in all 
respects or negative in all respects){\footnote{This is also clear in 
the treatments of balance in some other textbooks on SNA, for 
example the book of Scott \cite{S}.}}. The meat of his definition clearly 
concerns the set of ``all three such (pairwise mutual) relations'', not the 
pairwise relations themselves in isolation. Hence, though
Heider did not use the language of (di)graphs, it seems clear that he 
understood that balance could only be a useful notion if one
assumed symmetry. 
\\
\\
Now let $G$ be a graph on at least 3 nodes. We say that $G$ is {\em 
(completely) balanced} if every triad in $G$ is balanced. It is easy to see
that such a graph must either be a {\em clique} (all possible edges are
present) or a disjoint union of two cliques{\footnote{Here is a complete
proof of this fact, for the benefit of non-mathematical readers. Firstly, $G$ 
can have at most two connected components, because any triad whose three 
vertices all came from distinct components would be empty and hence 
unbalanced. Now let $x,y$ be two vertices in the same connected component.
We need to show that the edge $\{x,y\}$ is present in $G$. Since these
vertices lie in the same component, there must be {\em some} path between them,
say
\begin{eqnarray*}
v_0 := x - v_1 - v_2 - \cdots - v_k =: y.
\end{eqnarray*}
First consider the triad consisting of $x,v_1,v_2$. Two of three
edges are already present, namely $\{x,v_1\}$ and $\{v_1,v_2\}$. Since
all triads are balanced, the edge $\{x,v_2\}$ must also be present. Next
consider the triad formed by $x,v_2,v_3$. By the previous step, we already 
know that the two edges $\{x,v_2\}$ and $\{v_2,v_3\}$ are present.
Balance thus requires that $\{x,v_3\}$ also be present. We can keep iterating
this argument and deduce that $x$ is joined by an edge to every vertex $v_i$ 
along the path above, and hence finally to $y$.}}. As real-world (symmetric) 
social networks are rarely this simple, the notion of balance is not
very useful in SNA if taken literally. Indeed, its basic weakness lies in the
mottos M2-M4 which, in their informal expression, carry the 
assumption that the absence of a friendship implies its opposite, an emnity,
whereas in reality it may simply imply something like indifference. Hence, 
for example, a social network whose graph is a disjoint union of 3 or 
more cliques will not be balanced, since it will contain lots of
empty triads, even if the members of different cliques
merely have nothing in common and are not mutually antagonistic. Notice, 
however, that such a graph will still have no intransitive triads, which
supports the intuition that transitivity, as expressed by M1, is a much more 
coherent and fundamental idea than balance, as expressed by M1-M4. If a social
network is observed to possess a large number of intransitive triads, then
it indicates that something interesting is going on. This is the 
basic idea that will occupy us in the remaining sections of this paper. 
\\
\\    
A weaker, but potentially more useful, ``balance hypothesis'' would assert that,
in a real-life, symmetric social network, balanced triads should appear 
with greater frequency than in a graph of the same edge 
density where the edges are placed at random.   
Recall that, for a positive integer $n$ and a real number $p$ between 
zero and one, the Erd\H{o}s-Renyi random graph $G(n,p)$ is the random 
graph on $n$ nodes in which each of the $n(n-1)/2$ possible edges
appears with probability $p$, independently of all other edges. We can
now state the
\\
\\
{\sc General Balance Hypothesis (GBH):} Consider a social network in which all 
pairwise relationships are mutual, and hence the network can be represented as 
an undirected graph $G$. Suppose this graph has $n$ nodes and $e$ edges, thus 
edge density $p = \frac{2e}{n(n-1)}$. 
Let $i$ be either 1 or 3. Then the number of 
triads in $G$ in which exactly $i$ edges are present should exceed the expected
number of such configurations in the 
Erd\H{o}s-Renyi random graph $G(n,p)$. Similarly, if $i$ is either 0 or 2, 
then the number of 
triads in $G$ in which exactly $i$ edges are present should be less than 
the expected number of such configurations in $G(n,p)$.
\\
\\
If a network fails the balance hypothesis, in particular if intransitive 
triads are overrepresented compared to $G(n,p)$, then it is an indication that
something interesting is going on. For each $i \in \{0,1,2,3\}$, let
$E_i = E_i (n,p)$ denote the expected number of $i$-edge triads in 
$G(n,p)$, and $e_i = E_i/C(n,3)$ be the expected proportion of
such triads. Here $C(n,3) = \frac{n(n-1)(n-2)}{6}$ is the total number of
triads in a graph on $n$ nodes. We record the fact that
\begin{equation}
[e_0,e_1,e_2,e_3] = \left[(1-p)^3, 3p(1-p)^2, 3p^2 (1-p), p^3\right].
\end{equation}
The usefulness of GBH as a reference point is indicated by the fact that it is 
satisfied by 
the graphs considered above, which are disjoint unions of cliques. 
To prove this in full generality is a rather uninspiring calculus
exercise. For conceptual 
purposes, imagine the number $k$ of cliques as being fixed, suppose
the cliques have equal size $n$ and let the latter number tend to infinity. 
For large $n$, the edge density in the graph will be approximately $1/k$. 
Hence, by (3.1), the expected proportions of $i$-edge triads in the 
relevant Erd\H{o}s-Renyi graph will be approximately given by the vector
$\frac{1}{k^3}\left[ (k-1)^3, 3(k-1)^2, 3(k-1),1\right]$. By 
constrast, in the graph itself, one may check that
the corresponding proportions are approximately
$\frac{1}{k^3} \left[ k(k-1)(k-2), 3k(k-1), 0,k\right]$. Hence, 1- and 3-edge
triads are overrepresented, whereas 0- and 2-edge triads are underrepresented, 
in accordance with GBH. Of course, it is the complete absence of
intransitive triads which is the most striking feature.   
\\
\\
It is logically possible to extend the GBH to digraphs, in which case 
the assertion would be that balanced triads should be overrepresented
compared to a random digraph of the same edge density. 
However, such an extension of the hypothesis does not seem to add anything
conceptually. 
For, as we showed earlier, a balanced triad in a digraph must be
symmetric. If an experimenter, in constructing his network, decides
to make it directed, then he probably has a good reason for expecting
there to be a good deal of asymmetry. If it turns out that there is a bias
towards symmetry, at the level of dyads, then this bias will 
extend to any larger, symmetric configurations. 
Any additional bias towards balanced configurations should then be
interpreted, in the first place, with respect to the GBH for undirected graphs.
In other words, a balance hypothesis for digraphs is in essence nothing more
than the corresponding hypothesis for undirected graphs, together with 
a ``symmetry hypothesis'', which would assert that symmetric 
dyads should be overrepresented, in comparison to randomly constructed
digraphs. See Section 6 for some further discussion of the relevance of the 
latter.  
\par On the other hand, there may still be good reason to 
expect that transitivity, as expressed by M1, will usually be satisfied in 
directed networks in general. 
Property M1 seems reasonable in the absence of any 
assumptions about symmetry.  
Hence, for digraphs, it still seems useful to formulate a 
{\em transitivity hypothesis}. Note, though, that transitivity is a 
property, not of induced subgraphs (triads) but of ordered triples of nodes. 
We can
now state the
\\
\\
{\sc General Transitivity Hypothesis (GTH):} Consider a social network in 
which pairwise relationships are not necessarily 
mutual, and hence the network can be represented as 
a directed graph $G$. Suppose this graph has $n$ nodes and $e$ directed 
edges, thus 
directed edge density $p = \frac{e}{n(n-1)}$. 
Then the number of ordered triples $(x,y,z)$ of distinct nodes in $G$ which
don't satisfy M1 should be less than the expected
number of such triples in the 
Erd\H{o}s-Renyi random digraph $\vec{G}(n,p)$. Note that, in the latter,
each of the $n(n-1)$ possible directed edges is present, independently of the
others, with probability $p$. The expected number of triples 
not satisfying M1 is thus $n(n-1)(n-2)p^2 (1-p)$, since there are 
$n(n-1)(n-2)$ possible triples and for a triple
$(x,y,z)$ to fail M1, the directed edges $x \rightarrow y$ and 
$y \rightarrow z$ must both be present, while $x \rightarrow z$ is absent. The
first two events each occur with probability $p$ and the third with 
probability $1-p$. 
\\
\\
Let us now turn to the flawed treatment of these same concepts in 
Chapter 2 of \cite{Ka}. The problem begins with the author's apparent 
lack of understanding of 
transitivity. His first use of this term is on page 15, with the following 
sentence:
\\
\par ``{\em If the relationship is transitive, it means that if 1 loves 2, 
then 2 also loves 3}''.
\\
\\
Formally, he is saying the following:
\\
\\
M5. If $x,y,z$ are three distinct nodes in a digraph and if the directed edge 
$x \rightarrow y$ is present, then so is the directed edge $y \rightarrow z$. 
\\
\\
This is, obviously, not what transitivity means. In fact, the motto above is 
essentially meaningless, as the hypothesis concerns two entities, 1 and 2, 
whereas the conclusion concerns a third entity 3. There is no a priori 
relation between 3 and the others, he/she could be anybody. More formally, it 
is easy to prove that a digraph satisfying
M5 and containing at least four nodes{\footnote{If, in stating 
M5, we did not require $x,y$ and $z$ to be distinct, then we would have 
the same conclusion already for two nodes or more.}} must either be 
{\em complete}, i.e.: all 
pairwise directed edges are present, or {\em empty}, i.e.: all edges are 
absent{\footnote{Formally, if $n \geq 4$ then, modulo loops, 
there are only two possible relations on an $n$-element set 
satisfying M5, namely
the set of relations must either be empty or full. In contrast, for 
large $n$, it is known that there are close to $2^{n^{2}/4}$ transitive 
relations on an $n$-element, that is, relations satisfying the slightly 
stronger form of M1 where we don't require $x,y,z$ to be distinct. See
\cite{Kl}.}}
. The motto is therefore totally uninteresting.
\par Further down on page 15, the term ``transitive'' is used again, but now 
with the correct meaning. It then seems to be used properly for a while, until 
the end of Chapter 2, when on page 26 the original mistake is repeated in the 
following sentence:
\\
\par ``{\em Relationships are transitive when what holds for A to B, also 
holds for B to C}''.
\\
\\
The fact that the same incorrect statement is made in two different places 
is already quite worrying. 
This uncertainty regarding transitivity may be relevant to the 
extremely confusing analysis of ``balanced triads'' on page 25. Partly the
confusion arises from the author's failure to distinguish adequately between 
the notion of transitivity and the more restrictive notion of balance. 
More fundamentally, he doesn't seem to understand that a balanced triad must be
symmetric, and hence that the notion of balance is only really useful 
for undirected graphs, in other words for the analysis of social
networks in which there is an {\em a priori} reason to represent
relationships as being mutual. The high point of the confusion is when he gives 
triad types 7 and 8 (see Figure 1) as 
examples that ``conform to this hypothesis''. It's not entirely clear
if ``this'' refers to a transitivity or a balance hypothesis. But even if he
means the former then his assertion makes no sense. If he means
that these triads satisfy M1, then he is simply wrong, as the table on page 6
illustrates. If he means that, as digraphs, they satisfy GTH above, then
he is still wrong. Each of these digraphs contains 3 nodes and 3 directed 
edges, and 1 ordered triple of nodes failing M1. We compare with 
$\vec{G}(n,p)$ where $n=3$ and $p=3/6 = 1/2$. The expected number of
intransitive triples in the latter is thus $3\cdot 2 \cdot 1 \cdot \left( \frac{1}{2} \right)^3 = \frac{3}{4}$, which is less than 1, so both digraphs 
fail GTH.  
\par In my email correspondence with the author concerning Zachary's graph, it
became clear that he fundamentally misunderstood the concept of balance. It is
to these issues we turn in the next section.  

\setcounter{equation}{0}

\section{The karate club network of Zachary}

A classical study in the history of SNA was performed by Wayne Zachary, who 
observed the social interactions between members of a karate club over a 
period of approximately two years, from 1970 to 1972. He finally 
presented his results in 1977 \cite{Z} in the form of 
a graph (see Figure 4 at the end of the paper) 
showing the ``friendship'' connections between 34 club members near
the end of his observations and shortly 
before a formal split in the club. In 
other words, Zachary's graph had 34 nodes and each edge represented a pair of 
club members who were ``friends''. Crucially, Zachary assumed friendships were 
mutual, so his graph is undirected. It is also unweighted, though he also
considered a weighted version when considering information flow in the 
network{\footnote{Zachary ignored members of the karate club who did not 
interact socially at all. The club apparently had close to 60 regular members, 
hence a full representation of the social connections would have included up 
to 26 isolated nodes. One can make a strong case, I think, why it would have 
been better to include these nodes in the network. I will come back to this
point in Section 6.}}. 
The unweighted graph is reproduced on page 28 of \cite{Ka} and the author
then proceeds to perform a triad census. 
Recall that, in the usual mathematical terminology, a 
triad means an induced subgraph on three nodes. Hence, in an undirected graph, 
there are four possible 
types (i.e.: isomorphism classes) of triads, depending on whether the induced 
subgraph has 0,1,2 resp. 3 edges. 
\par On page 29, two main assertions are made, which we cite verbatim:
\\
\\
{\sc Assertion 1:} ``There are 1,575 symmetric dyads in the network (triad 
type 3-102 in chapter 2, figure 2) ... The number of dyads was much greater 
than would have been found by chance''. 
\\
\\
{\sc Assertion 2:} ``There are 45 (symmetric) triads in the entire network 
(triad type 16-300 in chapter 2, figure 2), also far more than expected by 
chance''. \\
\\
Unwinding the quantitative statements into standard mathematical terminology, 
the author is saying that the graph contains 1,575 triads in 
which one of the three edges is present, and 45 induced triangles. My own 
computer-aided check confirmed these numbers. 
However I also realised that the second part of the first assertion, that 
1-edge triads are overrepresented, is false, indeed very false. There are 78 
edges in 
this graph, out of a possible total of $C(34,2) = 561$. Hence, the appropriate 
comparison is with the Erd\H{o}s-Renyi random graph $G(n,p)$, where $n = 34$ 
and $p = 78/561$. By (3.1), the 
expected number of one-edge triads in the latter is
\begin{equation}
E_1 = C(n,3) \times 3p(1-p)^2 = \frac{n(n-1)(n-2)p(1-p)^2}{2} \approx 
1850.18 ...
\end{equation}
That the graph contains nearly 300 fewer one-edge triads seems significant - 
the probability of $G(n,p)$ containing so few such configurations is extremely 
small. Hence, Assertion 1 is false and the corrected version is as follows:
\\
\\
{\sc Assertion 1$^{\prime}$:} The number of one-edge triads in the karate club 
graph of Zachary is much less than would have been found by chance.
\\
\\
The expected number of 
induced triangles in $G(n,p)$ is 
\begin{equation}
E_3 = C(n,3) \times p^3 \approx 16.08 ...
\end{equation}
Hence Assertion 2 above is valid. After email consultation with the author it 
gradually became clear where his error with Assertion 1 
lay. He had computed expected values, 
not for $G(n,p)$, but instead for the directed version 
$\vec{G}(n,p)$. The configurations with which he was 
comparing the observed numbers of triads in Assertions 1 and 2 were, 
respectively,
\par - those in which one pair of directed edges was present, and all four 
other possible directed edges absent (triad type 3),
\par - those in which all six directed edges were present (triad type 16). 
\\
Let $\mathcal{E}_1$ and $\mathcal{E}_3$ respectively denote the expected 
numbers of these configurations in $\vec{G}(n,p)$. Then
\begin{equation}
\mathcal{E}_1 = C(n,3) \times 3p^2 (1-p)^4 \approx 190.68 ...
\end{equation}
and
\begin{equation}
\mathcal{E}_3 = C(n,3) \times p^6 \approx 0.04 ...
\end{equation}
These are consistent with the numbers the author showed me via email (the
numbers do not appear in the book), which he had 
obtained using a well-known software package called Pajek, in other words he 
did not use the exact formulas in (4.3) and (4.4). So it is clear where 
Assertion 1 came from. The conceptual mistake here is severe: it simply makes 
no sense to compare an undirected graph with random directed graphs. As the
equations above show, the resulting quantitative errors are enormous, and 
result in a qualitatively wrong conclusion, namely that the number of 1-edge 
triads is much larger than expected by chance, whereas in fact the complete 
opposite is true. 
\par It is clear that the author's reason for highlighting Assertions 1 and 2 
was to illustrate that the graph was well in accordance with the balance
hypothesis discussed in the previous section. Assertion 1$^{\prime}$ 
indicates that,
on the contrary, the evidence for this hypothesis is mixed: 3-edge triads
are indeed overrepresented, but 1-edge triads are significantly
underrepresented. To get a more complete picture, I also checked with 
a computer that the numbers of 0-edge and 2-edge triads in Zachary's graph are
3971 and 393 respectively. The corresponding expected numbers, $E_0$ and
$E_2$, in $G(n,p)$ are given by 
\begin{eqnarray}
E_0 = C(n,3) \times (1-p)^3 \approx 3818.95... \\
E_2 = C(n,3) \times 3p^2 (1-p) \approx 298.79.
\end{eqnarray}
Hence, both these types of triads are also overrepresented in Zachary's graph,
contrary to what the balance hypothesis would predict. In particular, the
overrepresentation of intransitive triads seems significant. Overall then, 
it is clear that Zachary's graph is highly ``unbalanced''.
\par After some email correspondence, the author admitted to me his conceptual 
and quantitative errors. However, he responded to my suggestion that
the unbalanced nature of Zachary's graph was an interesting phenomenon 
worthy of separate attention with the following message{\footnote{I realise that including details of email correspondence between two people puts the reader in the impossible position of being unable to directly verify the accuracy of what I write. I could have chosen not to mention my correspondence with the author at all, but then I would not have been able to acknowledge that he did at least admit his errors in the analysis of Zachary's graph. Having made this decision, I thought it best to give direct quotes, rather than my own interpretation of them.}}:
\\
\par ``{\em You are absolutely correct in one sense and wrong on balance in 
another sense. The graph is undirected and that is the only depiction of the
Karate club observations that make any sense. Hence the entire
discussion of a triad census and balance theory in this context is
incorrect since balance theory and the entire body of social network
theory that follows from it is only concerned with DIRECTED graphs.
Heider's original formulation was a directed graph (he did not have
those concepts then) discussion. Balance theory and its entire
literature therefore does not apply to undirected graphs.}''
\\
\\
I find these statements rather shocking since, as the previous section makes
clear, they demonstrate a complete misunderstanding of the underlying 
theoretical concept of balance. I will leave them to the reader to ponder, and 
instead turn to an investigation of the unbalanced nature of Zachary's graph.

\setcounter{equation}{0}

\section{A family of unbalanced graphs}

In this section, I will present a family of (random) graphs which exhibit the 
same pattern of imbalances in their triad counts as does Zachary's graph. 
In other words, in these graphs there are fewer 1-edge triads than in 
Erd\H{o}s-Renyi graphs of the same edge density, whereas all other 
triad types are overrepresented{\footnote{Since we shall be comparing two
infinite 
families of random graphs, all statements like this one should, if we are
being completely precise, be preceeded by words like ``almost surely as the
number of nodes goes to infinity ...''. To avoid getting too bogged down
in mathematical terminology, I will avoid uttering these words explicitly,
and leave it to mathematically inclined readers to fill in the gaps for
themselves.}}. This 
family will not exhibit all of the important structural features of Zachary's
graph but, I shall contend, is still rich enough to satisfactorily explain the
unbalanced triad census in the latter. Choosing a family with a simpler
structure will allow me to give rigorous proofs without becoming too
technical. We must also make an obvious caveat: Zachary's network is
just one specific graph, and here we shall be considering an infinite family
of random graphs. The reader should desist from taking any quantitive statements
made here and ``plugging in the numbers'' to Zachary's graph. Instead, the
graphs considered here are meant as idealisations, and are intended
to give a conceptual understanding of why Zachary's graph is
unbalanced in the way it is. 
\par For the remainder of this section, all graphs are assumed to be 
undirected. We begin with some standard mathematical terminology:
\\
\\
{\bf Definition 5.1.} Let $G$ be a graph on $n$ nodes. $G$ is called a 
{\em star graph} if it is a tree with $n-1$ leaves{\footnote{In graph theory, 
a {\em tree} is a connected graph with no cycles. A {\em leaf} in a tree is a 
node of degree 1.}}. 

\begin{figure*}[ht!]
  \includegraphics[]{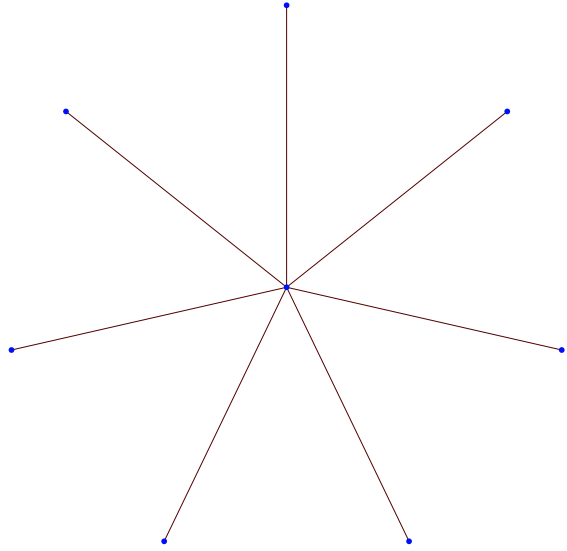} 
 \label{fig:ini}
\caption{A star graph with 7 leaves.}
\end{figure*}

Let $G$ be a star graph with nodes $v_1,...,v_n$ and suppose $v_2,...,v_n$ are
the leaves. Then $v_1$ is joined to every other node by an edge. We will
abuse terminology and refer to the node $v_1$ as the {\em star} in the tree.
Note that, in a star graph, there are no triads at all having either 1 or 3 
edges: the GBH could not fail more miserably. 
Suppose, however, that we now introduce what I think of as {\em random noise}. 
Precisely, let $\delta > 0$ be some small positive constant and, for each pair 
of leaves, insert an edge between them with probability $\delta$. We now have 
on our hands a random graph $G_{\delta}$, which I refer to as a 
{\em noisy star graph with noise parameter $\delta$}. The family of graphs 
which I will
now consider are disjoint unions of such random graphs. Here is the precise 
definition:
\\
\\
{\bf Definition 5.2.} Let $k,n$ be positive integers and $\delta \in (0,1)$ a 
(small) positive constant. For each $i = 1,...,k$, let $G_i$ be a noisy
star graph on $n$ nodes with noise parameter $\delta$. Let $G = G_{k,n,\delta}$ 
be the
disjoint union of the $G_i$, i.e.: the random graph whose connected components 
are the $G_i$. We shall refer to $G$ as a {\em $(k,n,\delta)$-noisy 
constellation}. 

\begin{figure*}[ht!]
  \includegraphics[]{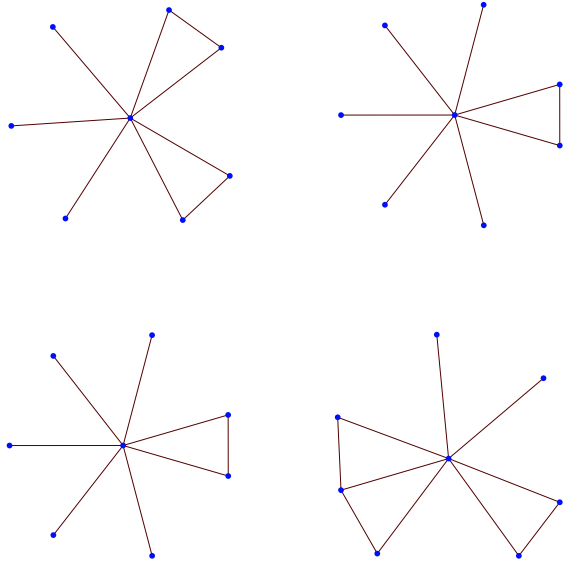} 
 \label{fig:ini}
\caption{A noisy 4-star constellation. Each of the noisy edges creates a triangle.}
\end{figure*}

$\;$ \\
The following standard notation will be used in the remainder of this section:
if \\
$f,g: \mathbb{N} \rightarrow \mathbb{R}$ are any two functions, we can
write either $f \ll g$ or $f = o(g)$ 
to denote that $\lim_{n \rightarrow \infty} f(n)/g(n) = 0$.
\\
\\
In what follows, we are interested in values of $k,n,\delta$ where
\begin{equation}
{\hbox{$k$ is fixed}}, \;\;\;\; n \rightarrow \infty, \;\;\;\;
\delta = \delta(n) = o_{n}(1),
\end{equation}
and all asymptotic estimates are to be interpreted with respect to
these conditions. 
\par The expected number of edges in a $(k,n,\delta)$-noisy constellation is 
given by 
\begin{equation}
\varepsilon = \varepsilon_{k,n,\delta} = k \left[ n-1 + \delta \cdot C(n-1,2) 
\right], 
\end{equation}
and the expected edge density is 
\begin{equation}
p = p_{k,n,\delta} = \frac{\varepsilon_{k,n,\delta}}{C(kn,2)} = \frac{\delta}{k}
+ \frac{2}{kn}(1+o_{n}(1)). 
\end{equation}
We wish to compare $G_{k,n,\delta}$ with the Erd\H{o}s-Renyi random graph 
$G(kn,p_{k,n,\delta})$. For each $i \in \{0,1,2,3\}$, let $\mathcal{E}_{i,a}$ 
denote the expected number of $i$-edge triads in $G_{k,n,\delta}$, and let 
$\mathcal{E}_{i,b}$ denote the coresponding quantity for $G(kn,p_{k,n,\delta})$. 
All of these quantities of course depend on $k,n$ and $\delta$, but we
suppress this in our notation, which otherwise would become unmanageable.
First consider $i=3$. Standard calculations yield
\begin{eqnarray}
\mathcal{E}_{3,a} = k \left[ \delta^3 \cdot C(n-1,3) + \delta \cdot C(n-1,2)
\right], \\
\mathcal{E}_{3,b} = p^3 \cdot C(kn,3).
\end{eqnarray}
If $\delta = o(n^{-1/2})$ then the second term in the expression for 
$\mathcal{E}_{3,a}$ dominates the first. By (5.3) it will also dominate the 
expression for $\mathcal{E}_{3,b}$ provided $n^{-2} = o(\delta)$. So
henceforth we shall assume that
\begin{equation}
n^{-2} \ll \delta \ll n^{-1/2}.
\end{equation}
In this range we will have
\begin{equation}
\mathcal{E}_{3,a} \sim \frac{k}{2} n^2 \delta, \;\;\;\;\;\;\;
\mathcal{E}_{3,b} \ll \mathcal{E}_{3,a}.
\end{equation}
Hence, 3-edge triads are likely to be {\em highly} 
overrepresented in $G_{k,n,\delta}$
as compared to $G(kn,p_{k,n,\delta})$. Next consider $i = 2$. Similar calculations
yield
\begin{eqnarray}
\mathcal{E}_{2,a} = k \left[ (1-\delta) \cdot C(n-1,2) + 3\delta^2 (1-\delta) 
\cdot C(n-1,3) \right], \\
\mathcal{E}_{2,b} = 3p^2 (1-p) \cdot C(kn,3).
\end{eqnarray}
Hence in the range (5.6) we will have
\begin{equation}
\mathcal{E}_{2,a} \sim \frac{k}{2} n^2, \;\;\;\;\;\;\;\;
\mathcal{E}_{2,b} \ll \mathcal{E}_{2,a}.
\end{equation}
Thus there will likely also be a large overrepresentation of 2-edge triads. 
Next consider $i = 1$. We have
\begin{flalign}
\mathcal{E}_{1,a} = k \cdot C(n-1,3) \cdot 3\delta(1-\delta)^2 + k(k-1)
\left[ n(n-1) + \delta \cdot n \cdot C(n-1,2) \right], \\
\mathcal{E}_{1,b} = C(kn,3) \cdot 3p(1-p)^2.
\end{flalign}
Here one has to work a little bit, but using (5.3) and (5.6) one can check that
\begin{equation}
\mathcal{E}_{1,b} - \mathcal{E}_{1,a} \sim k n^2.
\end{equation}
Hence, 1-edge triads are likely to be underrepresented in $G_{k,n,\delta}$, though
the difference from $G(kn,p_{k,n,\delta})$ will become less significant as 
$\delta$ increases beyond $n^{-1}$. More precisely,  
\begin{equation}
{\hbox{whenever $\delta \ll n^{-1}$}}, \;\;\;\;
\mathcal{E}_{1,a} \sim \left\{ \begin{array}{lr} k(k-1) n^2, & {\hbox{for
$k \geq 2$}}, \\ \frac{1}{2} n^3 \delta, & {\hbox{for $k=1$}}, \end{array} \right. 
\end{equation}
whereas
\begin{equation}
n^2 \ll \min \{\mathcal{E}_{1,a}, \mathcal{E}_{1,b} \}, \;\; {\hbox{whenever 
$n^{-1} \ll \delta$}}.
\end{equation}
The situation for 0-edge triads can now be deduced from our previous 
calculations. Since
\begin{equation}
\sum_{i=0}^{3} \mathcal{E}_{i,a} = \sum_{i=0}^{3} \mathcal{E}_{i,b} = C(kn,3),
\end{equation}
it follows from (5.7), (5.10) and (5.13) that
\begin{equation}
\mathcal{E}_{0,a} - \mathcal{E}_{0,b} \sim \frac{k}{2} n^2.
\end{equation}
Hence 0-edge triads are also overrepresented in $G_{k,n,\delta}$, though not
significantly since 
\begin{equation}
\mathcal{E}_{0,a} \sim \mathcal{E}_{0,b} \sim \frac{n^3}{6}, \;\; 
{\hbox{as soon as $\delta = o(1)$}}.
\end{equation} 
We can summarise our findings in a theorem, which we shall deliberately state 
somewhat informally:
\\
\\
{\bf Theorem 5.3.} {\em Let $G_{k,n,\delta}$ be a noisy constellation, where the
parameters $k,n,\delta$ satisfy (5.1) and (5.6). Let $p_{k,n,\delta}$ be as
in (5.3). Then for $i \in \{0,2,3\}$, the number of $i$-edge triads in 
$G_{k,n,\delta}$ is very likely to be significantly 
higher than in an Erd\H{o}s-Reny random 
graph $G(kn,p_{k,n,\delta})$. For $1$-edge triads, the opposite is true, though
their underrepresentation will be less significant once $n^{-1} \ll \delta$.
More precise quantitative statements are recorded in (5.7), (5.10), (5.13) and
(5.17) above.}
\\
\\
Note also that (5.7), (5.10), (5.14)-(5.15) and (5.18) imply that, 
for $k \geq 2$,
the expected number of $i$-edge triads in the noisy
constellations is a decreasing function of $i$ in the range (5.6).
For $k= 1$, the same is true once $n^{-1} \ll \delta$. 
\par In the next section we shall apply these findings to the analysis of
Zachary's graph.
 
\setcounter{equation}{0}

\section{Application to Zachary's graph}

The graphs considered in the previous section are models for social networks 
with the following characteristics:
\\
\\
(i) Pairwise relationships are a priori mutual, e.g.: friendships, so that
we have an undirected graph. 
\\
(ii) The network is split into a small number of groups of approximately equal
size. There is more or less 
no interaction between different groups, the reason for which
may depend on the particular network - in particular, 
the groups may be mutually antagonistic or just indifferent to one another.
\\
(iii) Each group is dominated by one individual, who is the ``star'' of his
respective group. This person maintains a relationship with every 
other member of his group.  
\\
(iv) Relationships between members of the same group, other than the star,
are generally weak. Some pairs of individuals do manage to form
a relationship, more or less at random. However, it is the 
relationships of the groups members to the star which are most
important. 
\\
\\
In Section 5 we demonstrated rigorously that, for a 
fixed number of groups of equal size, as the size of the groups increases and 
the
frequency of interactions between non-stars is not too large (see (5.6)), 
the triad census of such a network will reveal a significant overrepresentation
of 2- and 3-edge triads, compared to an Erd\H{o}s-Renyi random graph
with the same edge density. On the other hand, 1-edge triads will
be underrepresented, by an amount which becomes less
significant as the density of non-star interactions increases beyond an
intermediate threshold (see (5.15)). 0-edge
triads will be slightly overrepresented. The absolute numbers of $i$-edge
triads will be decreasing as $i$ goes from zero up to three (again, this
statement needs to be qualified if there is only one star - see the last
paragraph of Section 5). 
\\
\\
We saw in Section 4 that the triad census for Zachary's graph revealed 
the same patterns. And now we can see why, for the model in Section 5, 
with $k = 2$, is
clearly a reasonable idealisation of Zachary's graph. Shortly after he 
constructed his graph, showing the network of friendships between 34 club
members, the club formally split into two groups of 17 members each. Each
of these two groups had a star, the instructor Mr. Hi (node 1 in the
network) and the club president John A. (node 34), respectively. Indeed, before
the split Mr. Hi was friendly with 16 members, and all but one of these
joined his group afterwards. John A. was friendly with 17 people
beforehand and 15 of these joined his group. The remaining three people in the
network (nodes 17, 25 and 26) had a relationship with neither star
beforehand. Nobody joined a group unless they had a relationship 
with its star beforehand (in other words, all crossovers were friendly
with both stars beforehand). 
\par Still, Zachary's network is a bit more subtle than a 2-star 
constellation. The main reason for this is that 
there were three other ``minor stars'' who maintained a lot of
connections before the split. Node 2 had 9 friends, of whom 8 ended up in 
Mr. Hi's group. Node 33 had 11 friends, of whom 10 ended up in John A's
group. One gets the impression that nodes 2 and 33 acted as ``lieutenants''
for their respective stars in the ideological conflict preceeding the
split. Node 3, on the other hand, seems to have been the nearest the network 
had to a ``mediator''. He had 10 friends, of whom 6 ended up in Mr. Hi's group
and 4 in John A's. 
\par These five nodes (1,2,3,33 and 34) completely dominated the network. 
When one removes all the edges involving one of these five, then the remaining
network on 29 nodes contains only 19 edges, giving an edge density of
$19/C(29,2) \approx 0.047$, compared to an edge density of $78/561 \approx 
0.139$ for the network as a whole. Of these 19 edges, 9 were between members 
who both ended up in Mr. Hi's group and a further 9 were between members
who both ended up in John A's group. A solitary edge, $\{9,31\}$, connected
members who ended up on different sides and neither of whom were
stars or minor stars before the split.       
\\
\\
Hence, while the interactions in the karate club were certainly a bit more 
nuanced than in the toy model networks of Section 5, I think it is very 
reasonable
to assert that the latter capture the essence of what was going on in the 
club just before the split. What seems particularly significant here is the
weakness of the ties between ``ordinary'' club members (i.e.: 
non-stars and non-minor stars). Interactions between ordinary members who ended
up in different factions were almost non-existent (1 edge out of a possible
$14 \times 13 = 182$), but even those within each faction were weak (9 edges out
of a possible $C(14,2) = 91$ in Mr. Hi's faction, and 9 out of a possible
$C(15,2) = 105$ in John A's). In this situation, the fact that there
were approximately 26 club members who ``minded their own business'' and were 
not even included in the network analysis assumes greater significance. 
Had these been included, then the density of friendships between
ordinary members would have been a pitiful $19/C(55,2) \approx 0.013$. 
It is interesting, therefore, that on page 454
of \cite{Z}, Zachary writes the following:
\\
\par ``{\em Political crisis, then, also had the effect of strengthening the
friendship bonds within these ideological groups, and weakening the
bonds between them, by the pattern of selective reinforcement.}''
\\
\\
It is certainly very plausible
that the political conflict strengthened the ties of ordinary club
members to the various stars and minor stars, and may also have altered the 
strengths of
pre-existing friendships depending on the ideological adherence of the
people involved. Such things would be reflected more clearly 
in a weighted version of the
graph, something which Zachary indeed presented, but only at the same
fixed point in time so that it is not possible to see how the weighted network
evolved over time. However, I think the data hint at a more complex process. 
Consideration of the overall weakness of ties among ordinary 
club members, especially if
the 26 or so ``neutral'' members are included, suggests the following two
possible scenarios:
\par (i) in the absence of the ideological battle which 
served to focus members' attentions, the underlying network of 
friendships would have been very weak. Most members were
uninterested in socialising with others outside of karate lessons - they
generally did not regard a common interest in karate as a sufficient basis for
wider friendships.
\par (ii) the ideological battle actually served to stunt the development
of friendships between members who were not at the centre of the conflict, and
who began to see the club, not so much as a place to make friends, but as 
an ideological battleground where loyalty to one side or the other was the
main force driving interactions with other members.  
\\
Whatever the truth of the matter, it seems reasonable to consider the network 
drawn by Zachary, partly as a friendship network and partly 
as a network of loyalties in a split hierarchy.     
\\
\\
This brings us to more general sociological considerations on the notions of
transitivity and 
balance. Status differences seem to be a basic mechanism which mitigate 
against balance in configurations consisting of three entities or more. 
To see this, we first step back and consider two people, $A$ and $B$ say, 
interacting in 
isolation. Suppose $A$ likes $B$, but $B$, for whatever reason, is not 
interested in making friends with $A$. In terms of graphs, one imagines having 
a directed edge from $A$ to $B$, but no directed edge from $B$ to $A$. 
Intuitively, it seems clear that over time one of the following two things is 
likely to happen: (a) $A$ will succeed in winning over $B$ as his friend (b) 
$A$ will fail in getting $B$ to reciprocate his interest, and gradually lose 
interest in him, moving on to make other friends instead. In case (a), we will 
have two directed edges, in case (b) none. In case (a), we can 
replace the two directed edges by a single undirected edge. Hence, the 
following general claim seems reasonable in many situations{\footnote{Of course this claim will be false if the very basis of the relationship involves an
obvious asymmetry, for example employer-employee, leader-follower and so on. What we're interested in here is situations where the relationship is {\em a priori} symmetric, for example if it is based on some kind of homophily, so that a researcher's default hypothesis is that he is dealing with a network where the edges should be undirected.}} :
\\
\par ``{\em Pairwise relationships, considered in isolation, 
tend over time toward being mutual/symmetric.}''
\\
\\
The 
friendship between two people may be perfectly mutual as long as they have 
{\em something} in common, even if they are different characters in many
other respects. Suppose, however, that a third person enters the picture. Then 
the differences between the first two will affect the way they interact with 
the newcomer, which in turn will upset the mutuality of their own relationship.
Consider the following example: we have three people whom we call $A,B$ and 
$C$. $A$ plays football and also plays the piano. $B$ plays football but has 
no musical 
talent, whereas $C$ plays the piano but has no athletic ability. If $A$ and 
$B$ interact in isolation, then their common interest in football should lead 
to a ``perfectly mutual'' friendship, as they can simply ignore the other 
differences between them. The same applies to $A$ and $C$. But if all three 
interact together, then tension can arise from everyone's awareness of $A$'s 
higher ``status''. Both $B$ and $C$ are dependent on $A$ for friendship, as 
they have no basis for befriending one another. Hence, ``power'' becomes a 
factor 
in the relationships between $A$ and the others, which should be taken into 
account in any complete analysis of the social relations in the configuration 
as a whole. Indeed, over time, the relationship between $B$ and $C$ may move
from indifference to antagonism, as they compete for $A$'s attention. 
In the terminology of Section 3, the triad $ABC$ is intransitive, since two of 
three edges are present. What I think is most interesting, from a 
sociological/psychological viewpoint, is that tensions between $A,B$ and $C$
may not be evident if one just observes pairwise interactions in isolation.
People try to ``keep up appearances'' and maintain what look like
harmonious relations with their friends, while they simply
try to ignore people they may dislike. It is only by 
observing the intransitivity of the triad as a whole, especially if
it is part of a larger network in which such configurations are common,
that the observer might infer a lack of genuine mutuality at the level of
pairwise relationships.  
\par Note that, in the above example, the higher status of $A$ was a natural 
result of his wider range of talents. However, the same dynamic
could arise if $A$'s higher status was imposed from outside, i.e: if he
came to occupy a higher place in a wider social hierarchy. Suppose, for
example, that $A,B$ and $C$ are workmates, and that one day $A$ receives a 
promotion which places him in a managerial role above $B$ and $C$. Clearly,
this has the potential to fray all three pairwise relationships. However,
while $B$ and $C$ have the option, if worst comes to worst, of not
interacting at all, both must maintain some kind of relationship to $A$, he
being their boss. In this case, we'd still end up with an intransitive triad 
$ABC$,
with two of three edges present, but it would no longer be
appropriate to consider the edges as representing genuinely mutual
friendships, but rather as necessary interactions in an externally
imposed hierarchy. 
\par The above discussion considered intransitive triads only, but
we can extend it to understand how empty triads might come to be
overrepresented in a social network. If the network is
dominated a small number of high status individuals, then the dynamics
described above could stunt the development of friendships between
``ordinary'' network members, as they are drawn to, or compete for the 
attention of, the various stars. Hence, a lot of empty triads
involving ordinary members could arise.  
\\
\\
The relevance of these considerations to the karate club seems evident. 
On the one hand, 
recall that Zachary observed the interactions of the club members 
over a long time, more than 2 years. As we argued above, time seems to be of 
the essence in promoting mutuality in pairwise relationships, taken
in isolation. This supports the idea that 
Zachary was justified in assuming that friendships in the club were mutual 
and, hence, in making his graph undirected. Secondly, because the 
club is small, in a 2-year period every pair of members should have actually 
had the chance to meet and figure out 
whether they liked each other or not, so the absence of any particular edge in
the friendship graph cannot reasonably be attributed to the two parties
simply never having had a chance to interact. Thirdly, and most 
importantly, Zachary's decision to represent friendships as mutual is based
on his actual observations. We have no reason to doubt that this 
decision was reasonable, based on his observations of how
pairs in fact interacted. 
\par On the other hand, the club was racked by ideological conflict during
most of the period of observation. The two main figures occupied the 
central positions in the official club hierarchy, they being the instructor
and the president respectively. The data clearly suggest that, over time, 
it was the relationships of the club members to these two stars and their
respective lieutenants that drove the interactions in the club as a whole.
Friendships between ``ordinary'' club members were very rare overall.  
\par In particular, it is the overrepresentation of intransitive triads (393 as 
against an expected value in $G(n,p)$ of 299) that the above analysis picks
out as the most salient feature of the triad census in Zachary's network. 
This strongly hints at widespread 
tensions, even between members who were ostensibly friends, 
something which may not have been 
easy for Zachary to observe directly, as people tried to ``keep up
appearances''. Kadushin completely misses this 
point in his analysis, instead concentrating on 
the census of 1- and 3-edge triads, which he still manages to analyse 
incorrectly because of a serious conceptual error.  

\setcounter{equation}{0}

\section{Balance revisited}

In previous sections we have laboured to point out that the conventional
notion of balance, as expressed by M1-M4 in Section 3, is only really
useful to the social network analyst in situations where pairwise
relationships are {\em a priori} mutual, so that his default hypothesis is to
represent the network as an undirected, and unweighted, graph. To 
see this clearly, however, takes some mental effort, and the table on 
page 6 summarises the results of that effort. 
\par Suppose now, however, that we consider digraphs where loops are allowed, 
i.e.: directed edges of the form $x \rightarrow x$ from a node to itself. 
Mathematicians call such an object a {\em loop digraph}. 
Then M1-M4, in their formal expression, are still meaningful if we drop the 
restriction that the nodes $x,y,z$ must be distinct. Let 
M1$^{\prime}$-M4$^{\prime}$
denote the corresponding mottos, with this restriction removed. For a
mathematician, this is a natural step to take: let's see what it gives !
\par First consider a triple $(x,x,x)$, i.e.: the same node is
repeated three times. Then M2$^{\prime}$ implies that the edge $x \rightarrow x$ 
should be present. Hence, if a loop digraph is to satisfy M2$^{\prime}$, a loop
must be present at every node. This property is called {\em reflexivity}. 
Next consider a triple $(x,y,x)$, where $x \neq y$. We already know, by 
M2$^{\prime}$, 
that $x \rightarrow x$ is present. Suppose $x \rightarrow y$ is present. Then
M3$^{\prime}$ suggests that $y \rightarrow x$ should also be present. Conversely,
if we know $y \rightarrow x$ is present, then M4$^{\prime}$ 
suggests $x \rightarrow y$
should be so. In other words, if a loop digraph is to satisfy 
M2$^{\prime}$-M4$^{\prime}$, then
it must also be {\em symmetric}. 
\par To summarise, if we consider loop digraphs as the basic model for our
social networks, and formulate the notion of {\em balance} by 
M1$^{\prime}$-M4$^{\prime}$ instead,
then balance would automatically incorporate both reflexivity and 
symmetry{\footnote{In formal mathematical language, M1$^{\prime}$-M4$^{\prime}$ 
define a type of relation on the set of nodes in a loop digraph, which is both 
reflexive, symmetric and transitive, hence a so-called {\em equivalence 
relation}. In 
a completely balanced loop 
digraph, there can be at most two equivalence classes - 
see Section 3.}}. It's only a slight formal change in the definition, but it 
might help to avoid the kind of confusion which is evident in 
\cite{Ka} for example. In this context, we could also formulate a 
{\em General Balance Hypothesis for Loop Digraphs}, but this would now be a 
statement about ordered triples of nodes, rather than induced subgraphs on 
three nodes (triads). Such a hypothesis would assert that,
in certain kinds of social networks (the ``kinds'' being specified by
sociological criteria), the numbers of ordered triples $(x,y,z)$, of not 
necessarily distinct nodes, failing any of 
M1$^{\prime}$-M4$^{\prime}$ should be less than in 
a random loop digraph of the same edge density. Note that, in this setting,
if we have $n$ nodes and $e$ directed edges, then the edge density is
$p = e/n^2$, so that the expected numbers of triples failing 
M1$^{\prime}$-M4$^{\prime}$
in the corresponding random loop digraph are given, respectively, by 
\begin{equation}
{\hbox{Fail M1$^{\prime}$:}} n^3 p^2 (1-p), \;\;\;
{\hbox{Fail M2$^{\prime}$:}} n^3 (1-p)^3, \;\;\;
{\hbox{Fail M3$^{\prime}$:}} n^3 p^2 (1-p), \;\;\;
{\hbox{Fail M4$^{\prime}$:}} n^3 p^2 (1-p).
\end{equation}
One may ask why sociologists don't employ the notion of balance in this
modified form, but instead regard it specifically as a property of triads.
I am not a sociologist, so I cannot answer that question, but I will
hazard a guess, namely that it is because both reflexivity and
symmetry, taken on their own merits, are not {\em sociological} ideas about
{\em collectives}, but
rather purely {\em psychological} ones about {\em individuals}. First, consider
reflexivity. That a person maintain a friendly relationship with
himself seems like a basic psychological survival 
mechanism{\footnote{In everyday English, 
one can say that someone is ``unbalanced'', or that they are ``their own
worst enemy''. Both expressions roughly describe a person whose 
behaviour tends to do harm to themselves. This fits in well with the fact that,
as shown earlier, motto M2$^{\prime}$ implies reflexivity.}}. This driving force 
within
individuals also promotes symmetry between pairs. When faced with a 
choice between maintaining one's dignity and continuing a futile pursuit of
another's affections, a person will usually (though not always) choose 
the former option, especially given time. We also argued this point
in Section 6. 
\par Once three or more people are involved, however{\footnote{Indeed, a
reasonable definition of the word {\em society} is that it is any
collection of at least three people.}}, things can get a lot
more complicated. Some explicitly {\em social} factors,
such as status, can undermine balance, as we have discussed at length
in previous sections. Hence, in an intransitive triad, the two low-status
members may view their low relative status as a blow to their egos. On the
other hand, neither may be willing to let their jealousy of the other
jeopardize their friendship with the high status member. Even 
in a situation where two individuals share a deep mutual antipathy, there
may be a good reason for them to maintain a common friendship with a third 
person, especially if circumstances should one day force them to have
some dealings, since then their common friend can act as an effective 
go-between. Hence, in SNA, balance is a useful baseline concept, and the degree
to which a given network is balanced or not indicates the extent to 
which other, explicitly social factors, are at work.    
\par Even so, the notion of balance, in its
conventional usage, has serious limitations. It does not take account of the
fact that friendships or emnities can vary in strength - in particular,
it makes no distinction between emnity and simple indifference. It is
problematic to apply in large networks, where the absence of an edge
may be due to the fact that the two individuals involved never got a
chance to interact, alternatively to the fact that one or the other 
already has enough friends and simply has no time for any more. 
Underlying all this is the problem, 
stated repeatedly in this piece, that balance
is not a useful idea unless the pairwise social relationships are of a kind
that they should a priori be considered mutual. Expressing all this in
terms of graphs, we would want our graphs to be undirected, unweighted and
have a small number of nodes. 
\par Let us finish, therefore, by considering weighted digraphs in general.
There seems to an obvious, and useful, notion of ``balance'' in this
wider context, but it is quite different from the sociological notion. Namely,
one could say that a network is
``balanced'' if, at every node, the total weight of inward edges equals the
total weight of outward ones.  
Note that an undirected, unweighted graph is 
automatically ``balanced'' in this sense, but the converse need not hold. 
Indeed, an entire network may be ``balanced'' without any 
induced subgraph at all, on two or more nodes, having the same 
property. Triad type 10, consisting of a cycle of three directed edges,
is ``balanced'' in this sense, without 
being either symmetric or transitive. 
Hence, this notion of ``balance'' is totally
different from the sociological 
one, so much so that one really should use a different name{\footnote{There 
are, of course, only so many words in the English language, and sometimes
the same word is used to describe concepts which have nothing
whatsoever to do with one another. In pure mathematics, the 
word {\em balanced} is used about (undirected) graphs, but has nothing to do
with the number of edges in a triad. A graph is said to {\em balanced} if no
proper induced subgraph has a strictly higher ratio of edges 
to nodes. More precisely,
$G$ is balanced if, for every induced subgraph $H$ of $G$, one has
$\frac{e(H)}{v(H)} \leq \frac{e(G)}{v(G)}$.}}. 
The concept seems natural, though, and can be applied, for example, to
economic trading networks. In such a network,  
the weight of a directed edge $A \rightarrow B$ would represent the monetary 
value of all goods which $A$ sells to $B$. ``Balance'' then simply means 
that everyone is spending as much money as they are making. Of course, 
no real economic system, in particular any system which includes the
possibility of loaning money (a banking system), will ever be quite 
``balanced''. 

\setcounter{equation}{0} 

\section{Controversy}

As I explained in the introduction, the intial motivation for writing 
this piece came after reading the introductory sections of 
Charles Kadushin's recent textbook and realising just how flawed his thinking
was. I must admit I am rather baffled that nobody else seems to have yet 
made the criticisms outlined here. There are other books on SNA which
treat the same concepts with much greater care and accuracy, for example
Scott's book mentioned earlier \cite{S}.
Kadushin's book was published by 
Oxford University Press and has been formally reviewed by a 
number of experts in SNA. It seems to have been 
distributed widely among teachers and students.
Surely it should not have been left to a novice in the field
to point out its deficiencies ?

\section*{Acknowledgement}

I thank all my co-participants in an interdisciplinary reading course on 
Social Network Analysis currently being held at Chalmers University of 
Technology, without whom this note would never have made it into existence. 
In particular, 
I thank the course organisers, PhD students Vilhelm Verendel, Anton 
T\"{o}rnberg and Petter T\"{o}rnberg, who took the initiative to run the 
course and who selected the literature. I would also like to thank Magnus 
Goffeng for helpful comments on some earlier drafts of the paper.



\begin{figure*}[ht!]
  \includegraphics[]{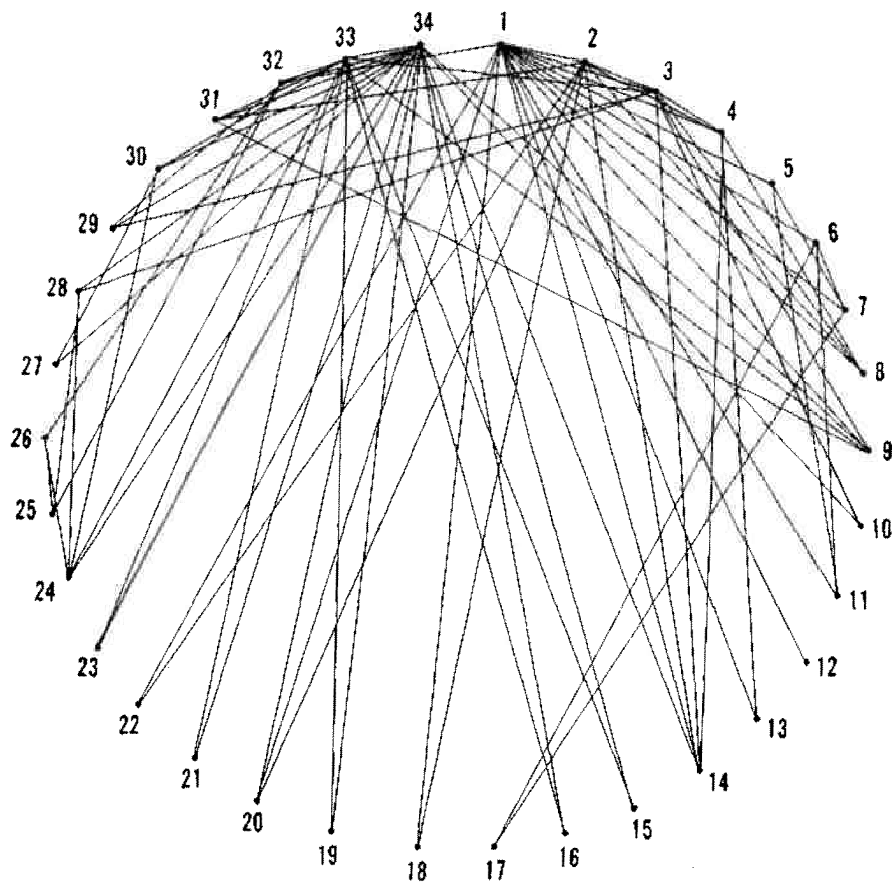} 
 \label{fig:ini}
\caption{Zachary's graph. In the graph on page 456 of \cite{Z} the edge $\{23,34\}$ is missing, but it is present in the matrix on page 457.}
\end{figure*}


\vspace*{1cm}

\begin{thebibliography}{ABRR} 



\bibitem[Ka]{Ka} C. Kadushin, \emph{Understanding Social Networks}, Oxford 
University Press (2012).

\bibitem[Kl]{Kl} J. Kla$\check{{\hbox{s}}}$ka, 
\emph{Transitivity and partial order},
Mathematica Bohemica \textbf{122} (1997), No.1, 75--82.

\bibitem[O]{O} \emph{The Online Encyclopedia of Integer Sequences}, 
Sequence $\#A000088$. \texttt{http://oeis.org/A000088}

\bibitem[S]{S} J. Scott, \emph{Social Network Analysis: A Handbook}, 2nd 
edition, SAGE Publications Ltd. (2000).

\bibitem[Z]{Z} W.W. Zachary, \emph{An information flow model for conflict and 
fission in small groups}, Journal of Anthropological Research \textbf{33} 
(1977), No.4, 452--473.
 
\end{thebibliography}
\end{document}